\documentclass[conference]{IEEEtran}
\IEEEoverridecommandlockouts
\usepackage{cite}
\usepackage{amsmath,amssymb,amsfonts}
\usepackage{algorithmic}
\usepackage{graphicx}
\usepackage{textcomp}
\usepackage{url}
\usepackage{wrapfig}
\usepackage{xcolor}
\usepackage{comment} 
\usepackage{enumitem}
 \usepackage[switch]{lineno}
 \setlist{leftmargin=5.5mm}
\newcommand{\bi}{\begin{itemize}}
\newcommand{\ei}{\end{itemize}}
\newcommand{\be}{\begin{enumerate}}
\newcommand{\ee}{\end{enumerate}}
\begin{document}
\title{  Fairer Software  Made Easier (using ``Keys'')}
\author{Tim Menzies,    Kewen Peng, Andre Lustosa\\
Computer Science,  NC State, USA\\
timm@ieee.org, kpeng@ncsu.edu, alustos@ncsu.edu}
\maketitle

\begin{abstract} 
Can we simplify  
explanations for software analytics? Maybe.
Recent results show that  systems often exhibit a ``keys effect'';
i.e.  a few key features control the rest.
Just to say the obvious, 
for    systems   controlled by   a few keys, 
explanation and control is just a   matter of running
a handful of  ``what-if'' queries across the keys.
By exploiting the keys effect, it should be possible to dramatically simplify even
complex explanations, such as those required for ethical AI systems. 
 
\end{abstract}
 \pagestyle{plain}
\section{Introduction}\label{intro}

Sometimes, AI must be explainable.
If some model  learned by an AI
 is to be used to persuade someone  to change what they are doing, it needs to be  explainable such that  humans can debate the merits of its conclusions.  
   Tan et al.~\cite{tan2016defining} argue that
        for software vendors, managers, developers and users, 
          explainable  insights  are the core deliverables of  analytics. 
        Sawyer et al. comment that explainable (and actionable) insights are the key driver for businesses 
        to invest in  analytics~\cite{sawyer2013bi}.    

Recently, the requirements for explainable AI have become quite complex. 
The European Union,
 Microsoft, and the IEEE have all released white papers discussing ethical AI~\cite{IEEEethics,EU,MicrosoftEthics}.   
While these  documents differ in the details,  they all agree that  ethical AI  must be ``FAT''; i.e.  fair, accountable and transparent.  Such ``FAT'' systems support    five ``FAT'' items:
       \be
       \item Integration with human agency;
        \item Accountability, where   old conclusions can be 
        challenged;
       \item Transparency of how conclusions are made;
       \item Oversight on what to change so as to change   conclusions;
       \item Inclusiveness, such that no specific segment of society is especially and unnecessarily privileged or discriminated against by the actions of the AI.
       \ee 
This list may seem like a   complex set
of requirements for explainable ethical AI.
However,
as seen in  this  paper,
software systems often have a ``keys effect''
by which {\bf a few key features control the rest}.
Just to say the obvious, 
when  systems are controlled by  a few keys, then
explanation and control is just a   matter of running
a few  ``what-if'' queries over the keys.
The rest of this paper discusses
(a)~how to implement the five ``FAT'' items; and
(b)~how   the keys effect simplifies all those implementations. 


\section{But First,  What is ``Explanation''?}\label{whatis}

Cognitive science argues that models comprising small rules are more explainable.
Larkin et al.~\cite{Larkin1335} characterize human expertise in terms of very small short term memory, or STM (used as a temporary scratch pad for current observation) and 
a very  large long term memory, or LTM.  
The LTM holds separate tiny  rule fragments
that explore the contents
of STM to say ``when you see THIS, do THAT''.
When an LTM rule triggers, its
consequence can rewrite STM contents which,
in turn, can trigger other rules. 
Short term  memory is very  small, perhaps even as small as  four to seven items~\cite{Mi56,Co01}.
Experts are experts, says Larkin et al.~\cite{Larkin1335} because the patterns in their  LTM
 dictate what to do, without needing to pause for reflection. Novices perform worse than experts,
says Larkin et al., when they fill  up  their STM with too many to-do's.  Since, experts post far fewer to-do's  in their STMs, they complete their tasks faster because (a) they are less encumbered by excessive reflection and (b) there is more space in their STM to reason about new information. 

First proposed in 1981, this   theory  still remains relevant\footnote{Recently,  Ma et al.~\cite{Ma14} used evidence from neuroscience and functional MRIs  to  argue  that STM capacity might be better measured using other factors than ``number of items''. But even they conceded that ``the concept of a limited (STM) has considerable explanatory power for behavioral data''.}. 
Humans best understand a model
which can ``fit'' it into their LTM; i.e., when that model comprises many small 
fragments.
Phillips et al.~\cite{phillips2017FFTrees} and others~\cite{gigerenzer2008heuristics,gigerenzer2011heuristic,czerlinski1999good} discuss how models containing tiny rule fragments can be  quickly comprehended by 
doctors in emergency rooms making rapid  decisions; or by soldiers on guard  making snap decisions about whether to fire or not on a potential enemy; or by  stockbrokers making instant decisions about buying or selling stock.
This theory-of-small-fragments explains both expert competency and incompetency in software
engineering tasks such as understanding code~\cite{Wi96}. 
Specifically,   expert-level comprehension means
having rules that   quickly lead to decisions, without clogging the STM.

It can be hard to find small transparent chunks of knowledge in
large data sets.
To some extent, FFTtrees~\cite{phillips2017FFTrees}, LIME~\cite{ribeiro2016should},
and SHAP~\cite{NIPS2017_7062} addresses that problem.
But a recent literature survey by Peng et al.~\cite{kewen21}
shows that   current    
explanation tools just rank the influence
of  features on single goals. Also,  LIME and SHAP  have  two
other limitations: they only show how single features effect single goals.
Also they rely on randomly generated instances. Such instance creation can overlook important structural details (such as the keys effect, described below). The explanations generated in that way might be over-elaborate, 
or irrelevant~\cite{kewen21}.

Hence, we seek a better framework that  
(a)~offers succinct explanations  from conjunctions of multiple
features on multiple goals; and which (b)~understands the structure of the data.



\section{What are ``Keys''?}\label{keys}

We have said in the introduction that the keys effect
means a few key features control the rest. Another way to say the say thing is
that most features in a system can be ignored or, more formally,
{\bf a system  with $N$ variables with $S$ values
usually visits much less than $S^N$ 
states}. This section
discusses this ``Druzdel effect''
and shows how it impacts the number
of key features that control a system.

\begin{wrapfigure}{r}{2in}
\begin{center}
 \includegraphics[width=2in]{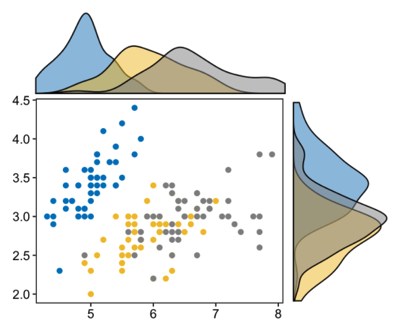}
 \end{center}
 \caption{Joint distributions  from software comprising two features running in three
 modes (blue, yellow, grey). }\label{joint}
\end{wrapfigure}

Imagine, in Figure~\ref{joint}, that  we
discretize the horizontal and vertical features into ten equal-width bins.
Once discretized in this way,   then  this data
has \mbox{$(S=10)^{N=2}=100$} states. But since the distribution of these features are skewed (particularly, the horizontal blue values), many of those  
states occur at zero frequency
(evidence: see   the white space in
Figure~\ref{joint}).

This Druzdel effect
(that most of the potential states within software are never visited)
has been observed in practice.
For example, Druzdel's~\cite{druzdzel2013properties} 
logged the frequency of observed states in  
a medical diagnostic system that monitored anesthesia patients in intensive care. 
Of the 525,312 observed states, one state occurred over half the time;
49 states covered 91\% of the probabilities; and the remaining states occurred vanishingly rarely
(at frequencies ranging from $10^{-6}$ to $10^{-22}$).

One explanation for this Druzdel effect is that when code runs, it only visits the $v$ states 
approved by the combination of its internal logic -- and this space can be far
smaller that $S^N$ (so $v\ll S^N$). 
Explanations, or controllers, of  software that visits  
\mbox{$v\ll S^N$} states only need to explore $\log_2{v}$ ``key'' features.
For example, Zhang et al. report that by
generating tests only for the main branches in the code, even applications that process large cloud databases can be tested via just a few dozen inputs~\cite{Zhang20}.

The number of   keys features in a data set can be inferred from
the number
of examples  required to model that data; e.g.
\bi
\item
 Druzdel found  $v_1=49$ commonly reached states.
\item
Yu et al.~\cite{zhe19} report that
security violations in 28,750 Mozilla functions 
can be
successfully modelled 
via an
SVM with \mbox{$v_2=271$} support vectors. 
\item Tu et al.~\cite{tu20} report that
defect labelling  in  $6000$ commits from Github
can be successfully modelled via a SVM with \mbox{$v_3=300$} support vectors~\cite{tu20}. 
\ei
Assuming binary features
and that the
number of  key features
is $N_i=log_2v_i$.
Hence,  the
above systems
could be controlled
via
$N_1,N_2,N_3\approx 6,9,9$
key binary features.

\section{Example}\label{eg}

This sections shows a small example
where keys find the tiny fragments of knowledge recommended by cognitive scientists.

Suppose an analyst wants to buy a fuel efficient lightweight car with good acceleration. The file
{\em auto93} (from the UC Irvine ML
  repository) has hundreds of examples of cars that mentions those features
as well as number of cylinders $c$, horsepower $h$, 
displacement $d$ (a.k.a. size of car), year of release $y$ and country of origin $o$.
From this data, we could apply  
(e.g.) linear regression   to learn
the model of Equation~\ref{one}:
 
{\scriptsize
\begin{equation}\label{one}
\begin{array}{r@{=}l} 
  \mathit{weight} &
 60.7{\times}c +
      5.2{\times}d +
      4.1{\times}h +
     13.7{\times}y +
    -48{\times}o +
    234.2 \\     
  \mathit{mpg} &
     -0.6{\times}c +
     -0.1{\times}d +
     -0.1h +
      0.6{\times}y +
      1.2{\times}o +
     -12.9 \\   
 \mathit{accel} &
      0.1{\times}c +
      0.1{\times}d +
     -0.1h +
      0.1{\times}y +
     -0.2{\times}o +
     21.2\end{array} \end{equation}
     }

This model is not tiny chunks of instantly interpretable  knowledge
recommended by cognitive scientists. In order to make recommendations about what kind
of car to buy, some further processing is required. For example, we could run a large number
of what-of queries across a large population of cars.
A standard tool for this purpose is a   genetic algorithm that 
(a)~mutates population of cars,   (b)~scores the mutants
on the above three equations, then (c)~selects the best scoring mutants as parents for
the next generation of mutants.
That genetic algorithm approach may not not scale  to large  problems.
Peng et al.~\cite{kewen21} has applied    genetic algorithms to explore $C=10^4$ software configuration options
(in order to optimize for $G=2$ goals). In that application,
the $O(GC^3)$ non-dominating sort procedure of the NSGA-II~\cite{deb02} genetic
algorithm took hours to terminate.  
A (much) faster approach, that generates tiny chunks of instantly interpretable knowledge,
comes from assuming the keys effect, as follows.
\bi
\item
If features control the output, then it follows that those
features, with different settings, appear in different outputs. 
\item
Also, if those features are few in number, then they will be a small set of features with most effect on the results.
\ei
To say that another way,
to exploit the keys effect. we only need to apply a little data mining.  
The follow three points describe a baseline keys-based algorithm called
KEYS0. The algorithm assumes that there exists a small number of feature ranges that occur with different frequencies in desired and undesired outcomes. Hence, a little random sampling is enough to 
quickly find those keys.
\be
\item   {\em Randomly divide the data.} Select a few random points to find two
distant points $P_1,P2$  within the data. Sort them such that point $P_1$ is better than $P_2$
(if exploring multiple goals, use some domination predicate to rank the two items).
Mark the data   {\em best, rest} depending on whether it is
closest to  
$P_1,P2$ respectively.
\item  {\em Look for
feature ranges
with very different frequencies in {\em} best
and {\em  rest}}. 
Divide numeric features into 
$\sqrt{n}$  sized ranges. Combine adjacent ranges that have similar {\em best,rest}
frequencies. Rank feature ranges, favoring those that are more frequent in {\em best} than {\em rest}. Print the top-ranked range. 
\item  {\em Recurse.} Select the data that matches that top  range (discarding the rest).
If anything  is discarded, loop back  to step1.
\ee
When applied to the {\em auto93} data, in 52ms, KEYS0
terminates in two loops after evaluating just four cars.
KEYS0
prints two tiny chunks of instantly interpretable knowledge;
i.e. the ranges ``cylinders $\le$ 4''
and ``origin==3''.

\begin{wrapfigure}{r}{2in}
    \includegraphics[width=2in]{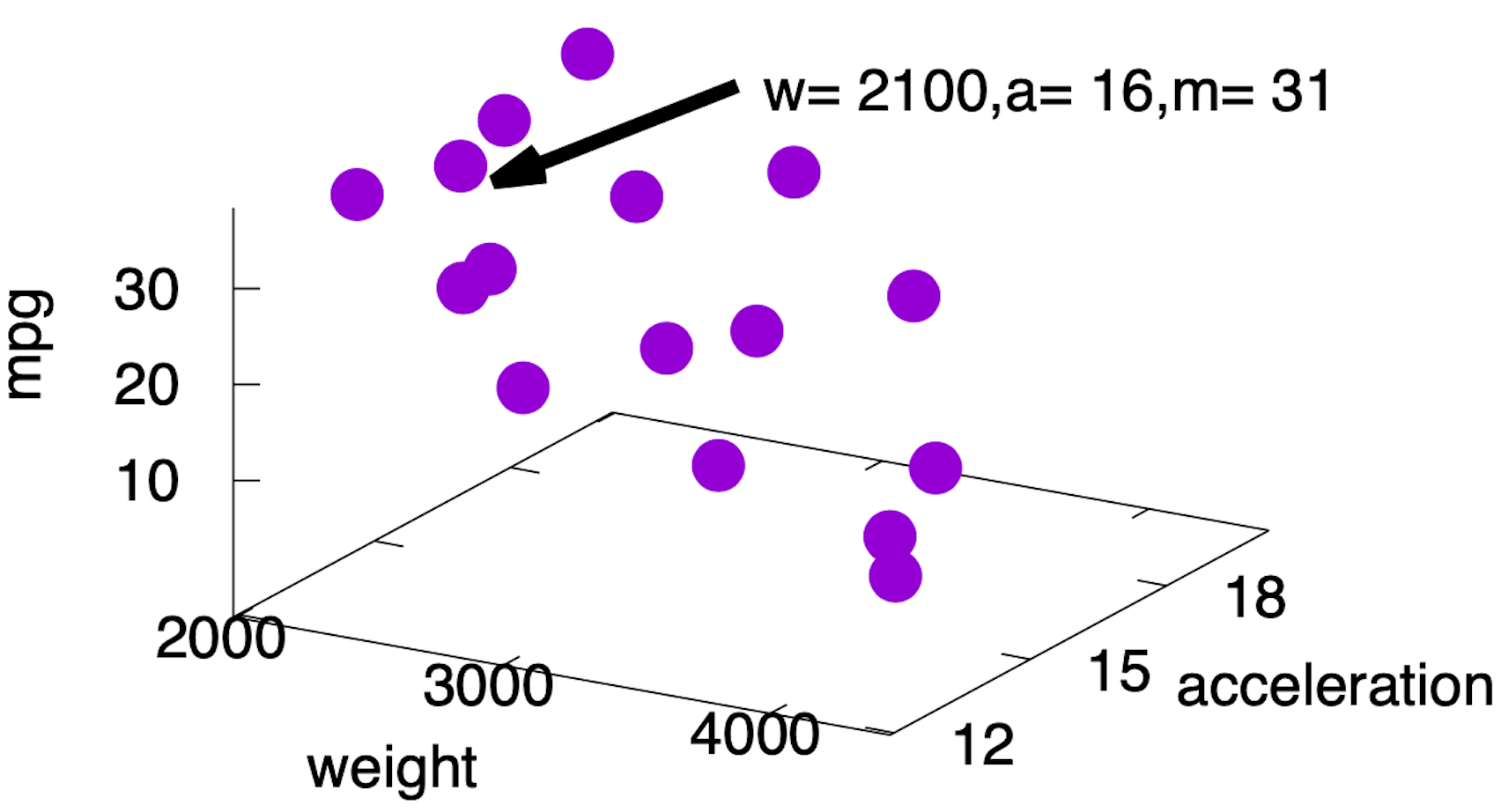}
    \caption{16 clusters of 398 examples from  {\em auto93}. Arrow
    shows  data found by KEYS0.}\label{auto}
\end{wrapfigure}

Figure~\ref{auto} shows the position of the data
selected by KEYS0.  In a result consistent with the presence of keys within 
{\em auto93}, we see that by examining just four examples andFairer Software Made Easier (using “Keys”) asking about two
ranges, we can find a useful
region in the data. Further improvements are
possible, of course (e.g. the  acceleration is still
mid-range in the final selected data). Nevertheless, the things to note here is (a)~how
little we had to probe the data in order to achieve such large improvements
via  tiny chunks of quickly interpretable  knowledge; and (b)~as far as we can tell,
the model
learned by KEYS0 is not apparent
within the output of standard learners
such as Equation~\ref{one}.

\section{What Causes Keys?}\label{cause}

Menzies~\cite{Menzies21}  speculates
that  keys arise from  {\em naturalness}~\cite{hindle16}
and/or {\em power-law}~\cite{lin15}  effects. Since programs exhibit the same frequency patterns  seen in human language,  then  {\em  naturally} we would  expect that code usually contains a small number of structures,  repeated many times~\cite{hindle16}.  As to {\em power laws}, suppose programmer2  most understands a small region  of the code written by programmer1.
That programmer  would tend to make  most changes around  that region.  If programmer3 does  the  same  for  programmer2's code,  and  programmer4 does the same for programmer3's code then that, over time, that team would  spend most
of
 their time working on a tiny portion of the overall code base~\cite{lin15}.
 In  such natural code that is  subject
 to  those  power laws, finding a controller for any part of the code  would mean that we also have a controller  for  many other  parts of the code (due to  repetition, due to the small
 size of  the  maintained code  base). 
 
 These effects are  not just
  theoretical speculation-- they  have  been seen in practice. 
 Lin and Whitehead report   power law effects
 they have observed in software~\cite{lin15}.
 Hindle et  al.~\cite{hindle16} report all the application  work in software analytics
 that successfully  exploits naturalness.
 Stallings~\cite{stallings09} comments that the design
 of operating systems often assumes
 a 
{\em  locality of  reference};
 i.e. that  program
 execution does not dash across  all the code
 all the time but instead is mostly
 focused  on a small part of
 the memory or instruction space.
 Lumpe et al.~\cite{lumpe2012learning} report  studies tracking decades of  work in dozens  of  open source projects. In  that code base,
 most classes are written once, revised never. Also, most  of  the  defect inducing
 activity is centered around a small number
 of rapidly changing classes. A similar pattern (where most activity is focused on a small part  of the code base)  has been reported in software from AT\&T~\cite{Ostrand04}, NASA~\cite{Hamill09},
 and within some widely used open source compiler tools (GCC~\cite{Hamill09}).


\section{Exploiting Keys for  Ethical AI}

Regardless of  their  cause,
this  section  argues
the keys effect can support the five ``FAT'' items
listed in the introduction.

\begin{wrapfigure}{r}{2in}
 \includegraphics[width=2in]{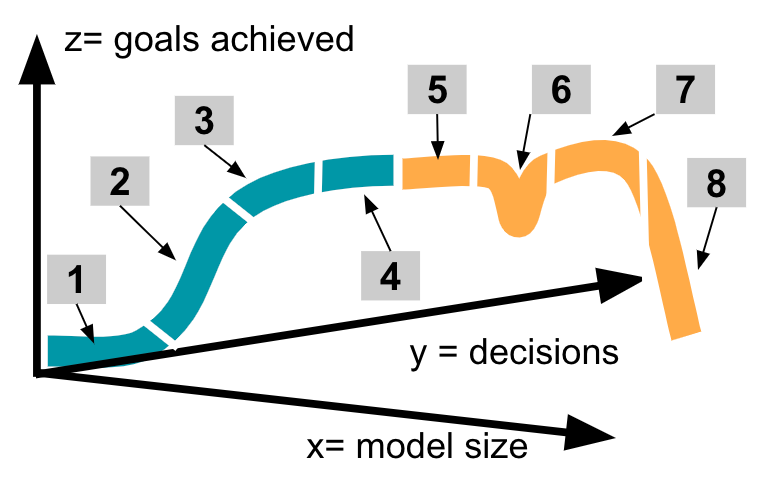}
 \caption{Model Building. (1) initial work; (2) some progress; (3) diminishing returns; (4) performance plateaus; (5) normal operation; (6) some problem; (7)  fixed;  (8)   collapse, after which the  cycle might start again.  }\label{intoai}
\end{wrapfigure}

{\bf Integration  with human agency:}
Figure~\ref{intoai}
shows   human+AI
interaction.
For  example,
{\em active}
{\em  learning} is where  AI and humans team up to solve problems. Whenever humans make a decision
about certain example, the active learner updates a model that predicts what humans might say about the next example.
Once the AI can adequately predict the human's opinion, then the human can retire and the AI can label the
rest of the data.
In Figure~\ref{intoai},
 one explanation system is better than another if  it leads to
higher {\em z = goals} are achieved with fewer $y=decisions$.

\begin{figure*}
\includegraphics[width=7.3in]{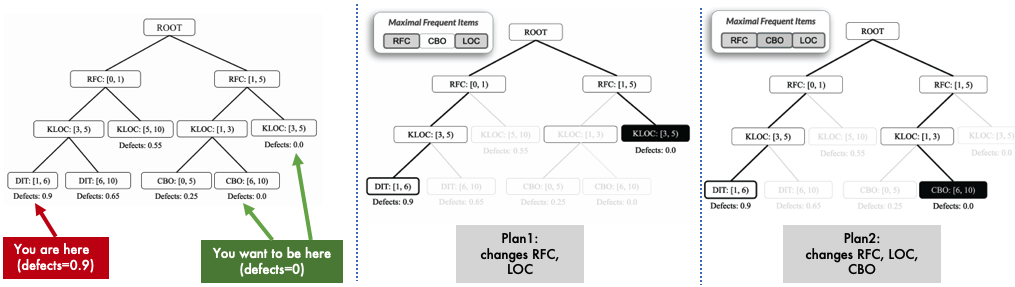}
\caption{Algorithm with oversight authority can request changes that alter conclusions. In this example,
the tree on the left has median number of defects per leaf. Using this structure, we can walk the tree
looking for  plans  that change the conclusion from defects = 90\% probable
to defects = 10\% problem (here, RFC, LOC, RBO are static code metrics that can be altered by different refactoring activities defined in Table~2
of~\cite{peng2021defect}). In this example, there is  historical data that RFC, LOC are never changed without also changing RBO.
Hence, we would propose the right-hand plan since it has most precedent in the historical record. Example from~\cite{krishna2020learning}.}\label{plan}
\end{figure*}
 We actually saw an example of Figure~\ref{intoai} in the above discussion.
Yu,Tu  et al.~\cite{zhe19,tu20}  incrementally updated an SVM model
based on an active learner. In a result consistent with
the keys effect, they needed only a few hundred support vectors
to model their data (Firefox security violations and the labelling of defect data
in Github).  We conjecture
that this result could be improved even further
(using fewer questions) by combining KEYS0 with the Yu,Tu tools.

{\bf Accountability:}
If we cache the y-axis decisions of Figure~\ref{intoai},
then the decision process that leads to the current system can be replayed by a third party. For systems demonstrating the keys effect,
then the total number of features that needs to be reviewed during an audit could be as low as a $\log_2{10^3}$ (see the calculations at the end
of \S\ref{keys}). While this might take days to weeks to study, such a manual audit is within the realms of possibilities by human beings.
Further, if we cache the entire rig that generated Figure~\ref{intoai}, then it would be possible to change some decision about the $y$ axis,
then replay the rest of the reasoning to see if that change effect the outcomes of the rest of the system.
Finally, if some part of the Figure~\ref{intoai} needs to be entirely revised then the keys effect promises that that revision
would not be an arduous process.

{\bf Transparency:}
 One issue with integrating with human agency is cognitive overload.
Humans and AI
tools can cope with different amounts of data. AI tools can readily process $10^8$   examples per minute, where example might contain
$10^2$  features, or even more~\cite{Witten:2002}. Human reasoning, on the other hand, is not so scalable.
While we cannot speak for all humans, this author asserts that reading   $10^8$ examples is just overwhelming.
Hence we say that  one explanation system is more transparent than another if it shows the user less data; i.e if it does
not  lead to cognitive overload.

Here again, keys can be useful.
Recall from \S\ref{eg} that we used keys to
explore hundreds of examples
with just two questions about two features.  In an even more impressive
example,  Lustosa et al.~\cite{lustosa21} recently used a variant
of KEYS0  to  explore different options for managing an agile SCRUM project.
That study explored 
10,000 examples with 128 features by asking the user $y \le 6$ questions (where each question mentioned only 4 ranges).  
Note that, once again,  needing to control so few features is yet another  example of how   keys  
can simplify explaining/controlling a system.

{\bf Human oversight:}
Krishna et al.~\cite{krishna2020learning} argues we can learn how to change conclusions using
 (a)~one    algorithm that makes predictions; then (b)~a second algorithm that   reports
the delta between regions with    different predictions.  
His variant of KEYS0 recursively divide the whole
data, without pruning any branches.
Next, he computed the   mean  conclusion seen 
in each leaf.
As shown in Figure~\ref{plan}, left-hand-side, we might prefer some leaf conclusions over the others (e.g. some leaves have a 90\% probability
of software having defects while other leaves predict a  0\% probability of defects). 
In that case, we can walk the tree from worse to better conclusions, each time querying the data division rule at each node.
If that process finds multiple ways to improve some conclusion then, as shown in Plan2 of Figure~\ref{plan}, we might prefer the plan
with most precedence in the historical record. In a result consistent with the keys effect, for data sets with 20 features,
the plans generated by this  procedure only needs to change 2-4 features~\cite{krishna2020learning,peng2021defect}.
 
 {\bf Inclusiveness:} there are many tests for checking if an AI tool is unfairly making conclusions at  (e.g.)  different false
 positive rates for different social groups. One way to repair an unfair learner  is to change how that learner builds its model via hyperparameter optimization~\cite{9286091,johnson2020fairkit}; i.e. planning  to change the learner control parameters such that the (say) false alarm rates are adjusted. In such optimization, the nodes of the trees in Figure~\ref{plan} would refer to control parameters of the learner while
 the leaves would be scored by (e.g.) the false alarm rates for the different social groups in the data.
With those changes, the planning process described in Figure~\ref{plan} could also be applied to unfairness reduction.

 \section{Discussion and Future Work}

We need to study explanation
 since explanation is the key
 to so much more; e.g., as shown above,
 controlling,  classification,  regression,  planning,  and optimization.
As part of that study, we should also
explore the keys effect since,
 as suggested
 above, those keys can simplify the generation
 of even complex types of explanations.

But keys does not solve all the problems of explanation and ethical AI. 
In fact, reviewing the above, we can see many open issues.
For example, do all data sets have keys? Is there some graduated explanation algorithm
that can shift easily from ``keys mode'' (for data sets with keys) to another mode
(for other kinds of data)?
KEYS0 currently divides data using a Euclidean distance metric.
But what about other dimensions as synthesised by (e.g.) some neural autoencoder?
Further,  what happens to Figure~\ref{joint} if humans make a wrong decision along the y-axis? How do we detect that mistake?
Furthermore, what happens when teams work together yet team members have conflicting goals?

The above paragraph is just a short sample
of the issues raised by our work
on keys-as-explanation
(for  other issues and research challenges raised buy this article, see    
 \textcolor{red}{{\bf \url{http://tiny.cc/todo21}}}).
In our experience,
exploring keys-for-explanation, we often find more research questions
than what we started with. 
But perhaps that is the true power of keys.
Keys-based explanation systems are so simple, so easily implemented, that
they encourage experimentation and, hence, the faster recognition of open
research issues. 
Our belief is that exploring those further issues will
be made easier by keys (and that is our current research direction).
Time will tell if that belief is correct.

\bibliographystyle{IEEEtran}
\bibliography{main}
 
 \clearpage
 
 \section*{Appendix: Research Directions}

  Looking though the above, we can see  numerous possibilities for future work. Some  issues
  relate directly to keys, while others
  relate more to  
  issues  with as
  human
  cognitive and, incremental
  knowledge acquisition.
  
{\bf RQ0) How to  evaluate ``explanation?''}  Many of the following will need a way to rank different explanation methods.
Hence, our first research question is how to certify an explanation evaluation rig.
If we say that a ``useful'' explanation means we know how to better build a  
model, then Figure~\ref{intoai}
  can be used for that ranking purpose.
  Specifically,
  explainer1 is better than explainer2 if more goals are achieved
  after making fewer decisions.
  
  For that evaluation, there are several cases.
  {\em Case1:} If the $y$ axis of 
  Figure~\ref{intoai} comes from some model, then the $y$ values
  can be scored via rerunning the model with new constraints
  learned from the explanation algorithm.   {\em Case2:}  If there is no model,
  and the $y$ values from some historical record of a project,
  then that data should be divided using time stamps into  
  {\em before} and  {\em after} data.   
The {\em after} data could be sorted according
  to its overlap with the recommendations generated from explanations built from the  {\em before} data. 
  Explainer1 is better than explainer2 if the after data
  with highest overlap is somehow better (e.g. fewer
  software defects) than the rest.

 {\bf RQ1) Can we build a keys
 detector?}
Research directions for such a detector including the following.
 Firstly, keys could be detected if,
 after applying feature and instance selection,
  the reduced data space generates models that perform as well
 as using all the data. For a discussion on this approach, see
 Papakroni, Peters et al.~\cite{peters2015lace2,papakroni2013data}.
 
 Secondly, run some very simple and very fast data miners run over the data building models using $N$ or $2N$ features.
 If any learner does no better than another, but  using only half the features,
 then kill the larger learner and start a new learner that uses
 half the features of the smaller learner. Keys would be defected if the final $N$ value is much smaller than the total number of
 features.  For a discussion on this approach, see~\cite{Holte93verysimple}.
 
 Thirdly,  researchers could (a)~take some well-explore SE domain;
 then (b)~commission the prior
 state-of-the-art (SOTA) method known for that domain; then (c)~compare results from (say) KEYS0 and the SOTA;  then (d)~declare that this domain
 has a keys effect if KEYS0 can control that domain using far fewer features than the SOTA.
 Note that, of the three approaches to {\bf RQ1} listed here, this third proposal would be the most labor intensive.
 
 {\bf RQ2) What is the generality of keys effect?}
 If we    have an automatic method for determining
 if a data set is amenable to keys, then we need to run that detector on as many data sets as possible in order to learn the extent
 of the keys effect.
 
 {\bf RQ3) Where do keys come from?}
 Another way  to assess the generality of the keys effect would be to understand what causes them. If we knew that,
 then we could apply the methods  of this paper whenever we see  those causative factors.  In \S\ref{cause},
 it was speculated that power laws within programmer teams lead to keys. To test that theory we could (e.g.)
 built defect predictors from projects that evolved from small single person into large teams. If power laws cause keys,
 then we would predict that algorithms like KEYS0 would be less/more effective earlier/later in that project life cycle (respectively).

 {\bf RQ4) How does keys compare to other explanation algorithms?}
 \S\ref{intro} and \S\ref{whatis} depreciated standard explanation algorithms like
  FFTtrees~\cite{phillips2017FFTrees}, LIME~\cite{ribeiro2016should},
and SHAP~\cite{NIPS2017_7062} arguing that they could only handle single goals and
(in the case of LIME and SHAP) only commented on
the effects of single variables on the goal. It is hence appropriate to check what is lost if we move from   multiple multiple variable multi-goal
reasoning (with KEYS0) to single variable and  single-goal reasoning (with the other approaches). If we re-run examples like \S\ref{eg},
and the net gain with KEYS0 over these other algorithms is minimal, and those other approaches
produced very small models, then that would challenge the motivation for this work.

{\bf RQ5)  Can keys simplify prior results?}
In the above, we speculated  that the
incremental active learning work of
Yu,Tu  et al.~\cite{zhe19,tu20}  could
be simplified  via keys. This  conjecture needs
to be checked.

Another example  of prior
work that might be  helped
by keys is incremental
{\em anomaly detection}.
Figure~\ref{intoai} proposed
an incremental knowledge capture
approach to system  construction.
In such an approach,  it is useful
to  have anomaly detectors
that understand when old  ideas
have  become out-dated, or when
new data is ``out-of-scope''
of the previously generated
model.    There are  many  ways
to  generate anomaly detectors~\cite{chandola2009anomaly}
and we  conjecture that a  succinct keys-based
summary  of past data might make it easier  to determine when new data is anomalous.  
 
 
 

{\bf RQ6) What are
the best algorithms for  finding  keys?}
The KEYS0 algorithm described above in \S\ref{eg} is very simple, possibly even simplistic.
Hence, it is worthwhile consider different ways to code that algorithm.

 The KEYS0 algorithm described above is a {\em greedy search} in that it prunes half the data at each level
 of its recursive search.  Lustosa et al.~\cite{lustosa21} prefers instead a global analysis where the whole
 cluster tree is generated, after which time the algorithm hunts around all nodes looking for things to prune
 and things to keep. Greedy algorithms run faster but global algorithms use more information about the data.
 Which works best? Or is there some way to amalgamate the two approaches (e.g. some local search in the style
 of MaxWalkSat~\cite{kautz96})?
 
Also, KEYS0 cluster the data via hierarchical random projects. There are so many other ways to cluster data 
(e.g. see \url{https://scikit-learn.org/stable/modules/clustering.html}) that it would be useful to explore other methods.
For example if the data is particularly complex, is there are any role for other
 algorithms such as 
 a neural autoencoder?  More generally,
 can we refactor 
 the code in steps1,2,3,4,5 into some object-oriented
 design with abstract super-classes like ``cluster''
and concrete sub-classes for
 different (e.g.) clustering algorithms? Such a refactoring
 would turn the above code into a workbench within
 which multiple algorithms could be mixed and matched
 in an effort to find better explanation algorithms.

No matter how the clusters are generated, some report must be made to the user
about the different between the clusters (this different was reported in \S\ref{eg} as the constraints 
 ``cylinders $\le$ 4'' and ``origin==''). KEYS0 uses a  simple greedy frequency-based method
 (combine  adjacent ranges that have similar {\em best,rest}
frequencies; tank feature ranges, favoring those that are more frequent in {\em best} than {\em rest}; print the top-ranked range)
and there are so many other feature and range ranking methods to try~\cite{ZHAO2021106652,hall2003benchmarking}
  
For simple data sets like the {\em auto93} data set explored in \S\ref{eg},
only 2 constraints are generated before the algorithm cannot find anything else of interest.
Lustosa et al.~\cite{lustosa21} reports studies in larger data sets (with 128 features)
where his keys finder returns very many constraints. Is there an early stopping rule
that can be applied to find enough useful constraints, but no more?
 
It is interesting to compare KEYS0 with genetic algorithms (GAs).
GAs mutates a {\em population} of, say, 100 individuals across many generations.
Each generation applies some select operator (to remove worse individuals) then combines the survivors to create the next
generation. KEYS0, on the other hand, applies its select opertor to  a much larger initial population, after which
it does no mutation (hence, no generation $i+1$). Prior results by Chen et al.~\cite{chen16} suggested that this over-generate-population0 can
do as well, or better, than multi-generation mutation (also, his approach is much faster than evolutionary methods). But evolutionary programming is a rapidly progressing field.
Hence, it would be wise to recheck Chen's conclusions against (e.g.) Hyperopt~\cite{Bergstra13}.
 
 {\bf RQ6) What is the impact of user mistakes?}
   One cognitive issue wit the  current KEYS systems is that  humans make many mistakes.
  What happens to Figure~\ref{joint} if humans
  make a wrong decision along the y-axis? How
  do we detect that mistake?
  For example,
  would some unsupervised outlier detection~\cite{Nam15} suffice? And how  do we repair those mistakes
  (Yu  et al. have some preliminary ideas on
  that~\cite{zhe19}).

  Another cognitive issue is that humans have different world views. Hence,
  reasoning about teams is very different to reasoning about one or two people.
  In the studies
 mentioned above regarding integration with
 human agency,
  Yu, Tu et al. worked mostly with one or two oracles~\cite{zhe19,tu20}. 
  How their methods scale to teams? Within the space
  of possible explanations, would we need to
  (e.g.) add in some multi-goal reasoning~\cite{harman2001search} so that
  we can appease the most number of competing goals
  as possible? On the other hand, rather than seek
  appeasement, it is better to report
  separate sets of conclusions where each set
  satisfies a different set of goals
  (e.g. see the Pareto clusters
  in Figure~5 of~\cite{veerappa2011understanding})?

 {\bf RQ7) Can visualizations
 replace rules?}  One of the more interesting results of 
  Lustosa et al.~\cite{lustosa21} is that large data spaces can be explored
  with reference to just a few ranges from a few features. Given that, can
  we offer users succinct visualizations of the data showing just the frequency counts of those ranges? And if users examined that space, would they find their own
  plans for controlling the data? And would those plans work better
  than 
  the plans generate via pure algorithmic means by Lustosa et al.~\cite{lustosa21}?

 {\bf RQ8) How  to  maintain
 privacy and accountability?} In the above,
 it was proposed to cache the rig of  Figure~\ref{joint}, then re-run it
for {\em accountability} purposes. This raises two data management systems
issues. 
Firstly, if we retain some or all
the data of Figure~\ref{joint}, do we need some (e.g.) mutation policy to reduce
the probability of identify disclosure within that data?~\cite{peters2015lace2}?
Secondly,  how much past data do we need to keep in order to support
future accountability? If we must keep {\em all} the past data then that could
become a significant data storage issue. On the other hand, recalling Steps 1,2,3,4,5,
would it suffice to just  keep the distant points $P_1,P_2$ found at each level
of the recursion? Or is the best data retention policy somewhere in-between?
(Aside: see Nair and Peters et al.~\cite{nair2018faster,peters2015lace2} for different proposals
on  selecting the most
informative prior data).

\end{document}